\documentclass[11pt]{article}
\usepackage{moriond,epsfig,amssymb,amsmath}
\usepackage[rflt]{floatflt}

\newcommand{\KL}{\mbox{$\kl$~}}
\newcommand{\KS}{\mbox{$\ks$~}}

\newcommand{\KZ}{\mbox{$\kz$~}}

\newcommand{\reoe}{{\mathrm Re}(\epsi'/\epsi)}
\newcommand{\EPOE}{\mbox{$\epsi'/\epsi$}~}
\newcommand{\REOE}{\mbox{{\rm Re}($\epsi'/\epsi$)}~}
\newcommand{\epsi}{\varepsilon}

\newcommand{\kl}{{\rm K_{L}}}

\newcommand{\ks}{{\rm K_{S}}}

\newcommand{\kz}{\rm K^{0}}
\newcommand{\ETA}{\mbox{$\rm \eta$}~}
\newcommand{\pz}{{\rm\pi^{0}}}
\newcommand{\pppm}{{\rm\pi^{+}\pi^{-}}}
\newcommand{\pzpz}{{\rm\pi^{0}\pi^{0}}}
\newcommand{\epsfigure}[5]{\begin{figure}[#1]\centerline{\epsfig{file=#2,width=#3\textwidth}}\caption{#4}\label{#5}\end{figure}}

\begin{document}
\vspace*{4cm}

\title{LATEST RESULTS ON KAON PHYSICS FROM THE NA48 EXPERIMENT}

\author{ GIACOMO GRAZIANI }

\address{Laboratoire de l'Acc\'el\'erateur Lin\'eaire, Orsay\\
        {\rm on behalf of the NA48 collaboration}}

\maketitle\abstracts{ The NA48 experiment, conceived primarily to look 
for direct CP violation in neutral kaon decays, has recently published
the so far most precise determination of the \EPOE parameter. After
reviewing shortly this result, we report on the 2001 data--taking,
which concluded the $\epsi'$ program by collecting a substantial
amount of data with different beam intensity conditions. 
We also present new precision measurements of the \KZ and \ETA masses and
of the \KS lifetime, that provide consistency checks of our
analysis. Finally, the prospects for the future experimental program
are discussed.
}

\section{The measurement of \REOE}
The CP--violating two--pion decay of the long--lived neutral kaon, 
dominated by its CP=+1 component $\rm K_1$, can also proceed directly in the 
decay \mbox{$\rm K_2\to\pi\pi$} through the interference of the
\mbox{$\kz\to\pi\pi$} amplitudes $A_I$ with isospin $I$=0
or 2. This direct CP violation is usually parametrized through  
the quantity
\begin{equation}
 \epsi' ~=~ \frac{i}{\sqrt{2}}
Im\left(\frac{A_2}{A_0}\right)
e^{i(\delta_2-\delta_0)} ~~~ \left(\text{phase convention: } Im(A_0) \equiv 0\right)
\end{equation}
In the Standard Model (SM) picture, \EPOE is
proportional to the CKM parameter $Im(\lambda_t)$ and is expected to
be of the order $10^{-3}$, though the uncertainty in the calculation is
dominated by long distance hadronic contributions (see \cite{soni} for 
a review). Nevertheless, a high--precision measurement of \EPOE can
test the SM prediction against other possibilities, as the Superweak
Model (predicting $\epsi'=0$) or large contributions from new physics.
All experiments performed so far have measured \REOE through the double ratio
method~\footnote{being the phase of
$\epsi'$ accidentally very close to that of $\epsi$ ($\simeq -\pi/4$),
we get \EPOE$\simeq$\REOE}:
$$ {\mathrm R} = \frac{\Gamma(K^0_L\to\pi^0\pi^0)}{\Gamma(K^0_L\to\pi^+\pi^-)}
      \frac{\Gamma(K^0_S\to\pi^+\pi^-)}{\Gamma(K^0_S\to\pi^0\pi^0)} ~\simeq~
1-6\times {\rm Re}\left(\frac{\epsilon'}{\epsilon}\right)  $$
exploiting the cancellation of many experimental uncertainties in the
ratio. The two dedicated experiments performed during the eighties (NA31
at CERN~\cite{na31final} and E731 at Fermilab~\cite{e731}) did not reach 
a definitive conclusion about the occurrence of direct CP violation
and a second generation of experiments was needed, which eventually provided a
convincing evidence for a non--zero effect after their latest results
were announced during 2001:
\begin{tabbing}
~~~~~~~~~~~~~~
\=NA48 at CERN SPS~\cite{epsi01}:~~~~~~~~~~~~~~~~~~~\=$\reoe = (15.3 \ \pm \ 2.6) \times 10^{-4}$ \\
\>KTEV at FNAL (preliminary)~\cite{graham}:	\>$\reoe = (20.7 \ \pm \ 2.8) \times 10^{-4}$ 
\end{tabbing}

\medskip
The method used by NA48 consists in measuring the four decay modes
simultaneously from the same fiducial region using two high--intensity
and quasi--collinear \KS and \KL beams (see figure \ref{beams}). The
two beams illuminate in a very similar way the central detector, based 
on a large magnetic spectrometer and on a liquid Krypton (LKr)
homogeneous calorimeter, where the $\pppm$ and $\pz\pz$ decays are
reconstructed~\footnote{for a more detailed description of beam
lines and detectors see (~\cite{epsi01})}.
\epsfigure{tb}{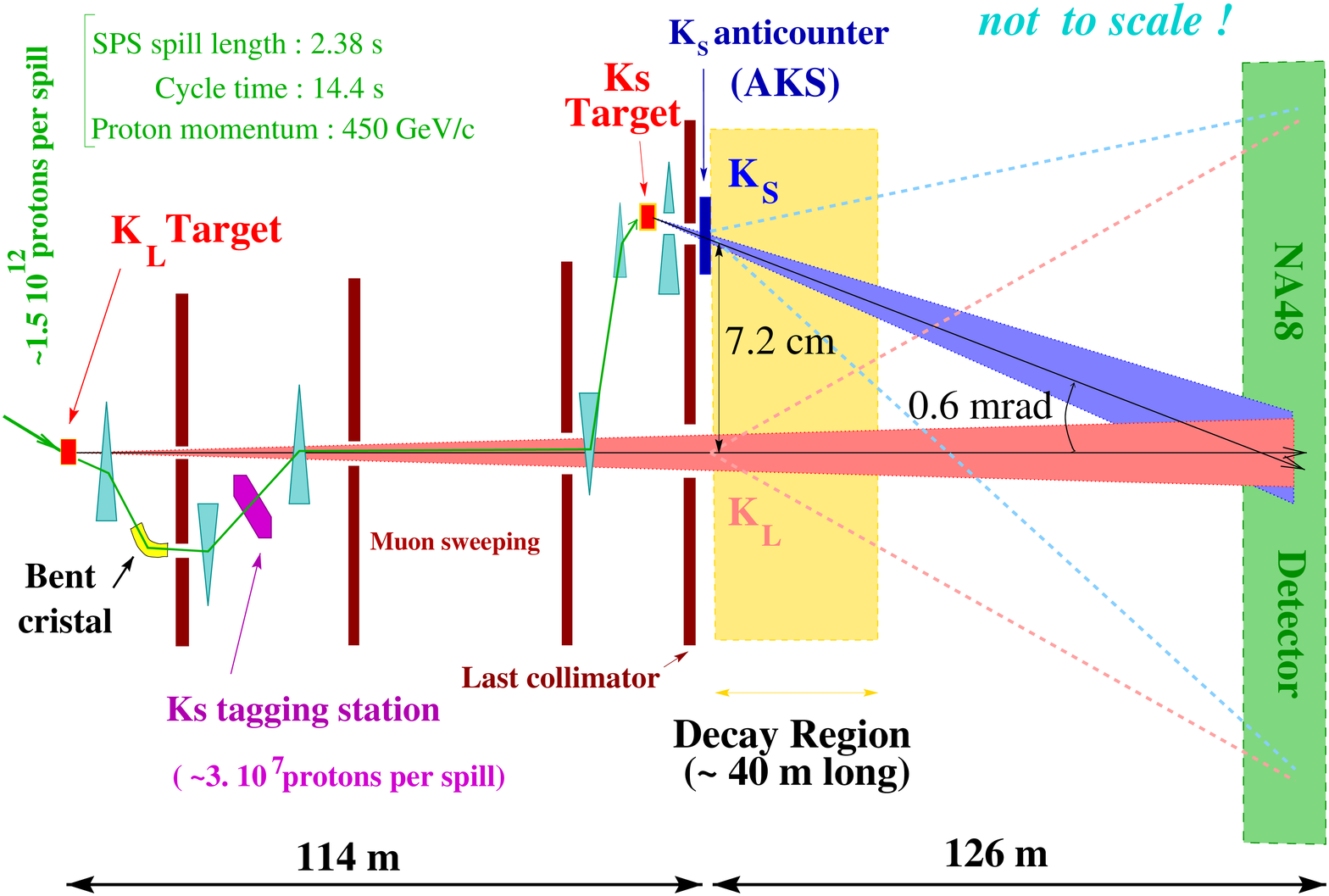}{.62}
{A sketch of the simultaneous \KL and \KS beam lines. The fiducial
region for \KZ decays 
begins more than 12 \KS lifetimes downstream the primary target in
order to obtain an almost pure \KL beam. The \KS beam is obtained by
extracting a small fraction of the protons emerging from the first
target and transporting them to a second target located immediately
before the decay region. The two beams converge with a small angle in
order to be overlapped in the middle of the central detector. Numbers
in the figure refer to the 1998/1999 configuration.}
{beams}
In order to distinguish \KS
from \KL events, a $\pm$ 2 ns coincidence is required between the
event time and the passage of a proton in a tagging station located
along the \KS beam line.
The two main differences between \KS and \KL are minimized offline:
\begin{itemize}
\item the analysis is performed in 20 kaon energy bins between 70 and 170
GeV to account for the different energy spectra;
\item \KL events are weighted according to the \KS lifetime to
equalize the effective detector illumination from the two beams.
\end{itemize}
Finally, a set of small ($<$0.3 \% by first principles) corrections have 
to be applied to account for remaining biases as residual acceptance
difference, backgrounds, $\rm K_L\leftrightarrow K_S$ mistagging,
intensity and reconstruction effects. 

\section{The 2001 run}
The result recently published~\cite{epsi01} by NA48 has been obtained
from the data collected during the 1998 and 1999 runs, corresponding
to $3.3\times10^{6}$ $\rm K_L\to\pzpz$ (the decay mode
limiting the statistical accuracy). The 2000 run was used to perform
some cross-checks and other physics measurements on neutral decay
modes, the spectrometer being unavailable after that all its four drift 
chambers were damaged in an accident occurred in November
1999. Meanwhile, the chambers were rebuilt and reinstalled in time for
the 2001 data--taking. 

\begin{table} \begin{center}
\caption{Corrections and errors on the double ratio for
the 1998+1999 data, listed in decreasing uncertainty.}
\label{9899result}
\begin{tabular}{|l|rrr|}  
\hline
Statistical error	      & --   & $\pm$ 0.00101  & \\
\hline
$\pzpz$ reconstruction        & --   & $\pm$ 0.00058  & \\
Acceptance            & +0.00267   & $\pm$ 0.00057    & \\
$\pppm$ trigger inefficiency  & --0.00036 & $\pm$ 0.00052 & $\leftarrow$ rate effects\\
Accidental activity           & --  & $\pm$ 0.00044  &  $\leftarrow$ rate effects\\ 
Accidental tagging    & +0.00083 & $\pm$ 0.00034   &  $\leftarrow$ rate effects\\
Tagging inefficiency  & --   & $\pm$ 0.00030      &  $\leftarrow$ rate effects\\
Background to $\pppm$ & +0.00169  & $\pm$ 0.00030     & \\
$\pppm$ reconstruction        & +0.00020   & $\pm$ 0.00028   & \\
Beam scattering       & --0.00096 & $\pm$ 0.00020     & \\
Background to $\pzpz$ &--0.00059 &  $\pm$ 0.00020     & \\
Long term \KS/KL variations   & --   & $\pm$ 0.00006  & \\
\KS anticounter inefficiency      & +0.00011  & $\pm$ 0.00004     & \\
\hline 
Total systematic       &   +0.00359 & $\pm$ 0.000126  & \\ 
  \hline 
\end{tabular} 
\end{center}
\end{table}

The systematic corrections on the double ratio for the 1998/1999 analysis are
listed in table \ref{9899result}. Several 
sources of error are related to rate effects, namely the residual
differences of instantaneous intensity seen by \KS and \KL events
(leading to possible differences in accidental activity and trigger
inefficiency) or by neutral and charged events (leading to possible
differences in the mistagging probabilities, which depend from the \KS
proton rate seen by the reconstructed events). For this reason the
2001 data were taken with different beam conditions: profiting of 
the possibility to extend the SPS duty cycle after the closure of LEP
(5.2/16.8 instead of 2.4/14.4 s), we could decrease the average
instantaneous intensity by about 30 \% while keeping about the same typical
per day event statistics (see figure \ref{2001beam}). Concurrently the
proton energy was decreased from 450 to 400 GeV.   
\epsfigure{htb}{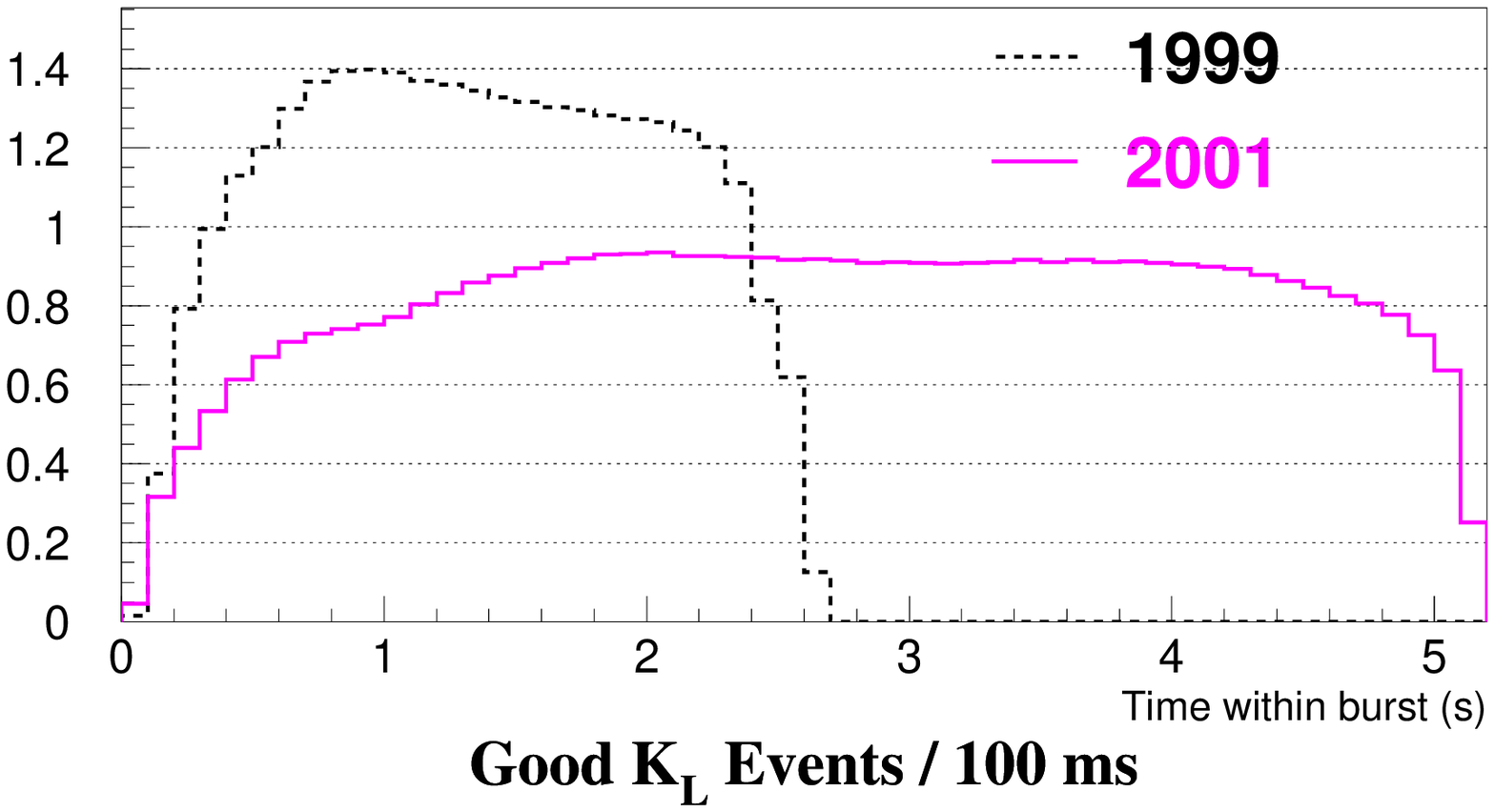}{.5}
{Rate of selected \KL events along the proton burst for the
2001 and 1999 runs}
{2001beam}

The data--taking was successful, with more than $1.4\times 10^{6}$
$\rm K_L\to\pzpz$ recorded, corresponding to about
$14\times 10^{-4}$ statistical error on $R$. This should allow to decrease the
final statistical error on \REOE from $1.7\times 10^{-4}$ to $1.4\times
10^{-4}$. The performances of the new spectrometer were very similar
to the previous runs (about 2.5 \mbox{MeV/c$^2$} resolution on the
$\pppm$ invariant mass). All the 
effects related to intensity were reduced as expected: for example,
the efficiency of the level--2 charged trigger increased from 98.3\% to 99.2\%
and the probability of an accidental coincidence between a \KL event and a
\KS proton was reduced from 10.6\% to 8.1\%.

We expect that the total error (statistical plus systematic) on the
double ratio from these data will be comparable to the published
result, so that the 2001 run will be a major cross-check of the \REOE
measurement against intensity effects.

\section{Neutral energy scale and the masses of \KZ and \ETA}
The longitudinal decay position for \mbox{$\rm K^0\to\pzpz\to
4\gamma$} is reconstructed by imposing the \KZ mass to the four
detected photons, whose transverse positions and energies are precisely
measured by mean of the LKr calorimeter. Thus, the global calorimeter
energy scale fixes the longitudinal distance scale for the neutral mode,
that must be identical to the scale for the charged mode (fixed by the
spectrometer geometry). 
In order not to be too sensitive to the uncertainty on
energy scale, the beginning of the fiducial region is defined by an
anticounter on the \KS beam line. The energy scale is fixed by
adjusting the anticounter  position reconstructed from the decay
vertex distribution (see fig. \ref{aks_decay_neutral}) to its known
true position.  
\begin{floatingfigure}[htb]{.45\textwidth}
\centerline{\epsfig{file=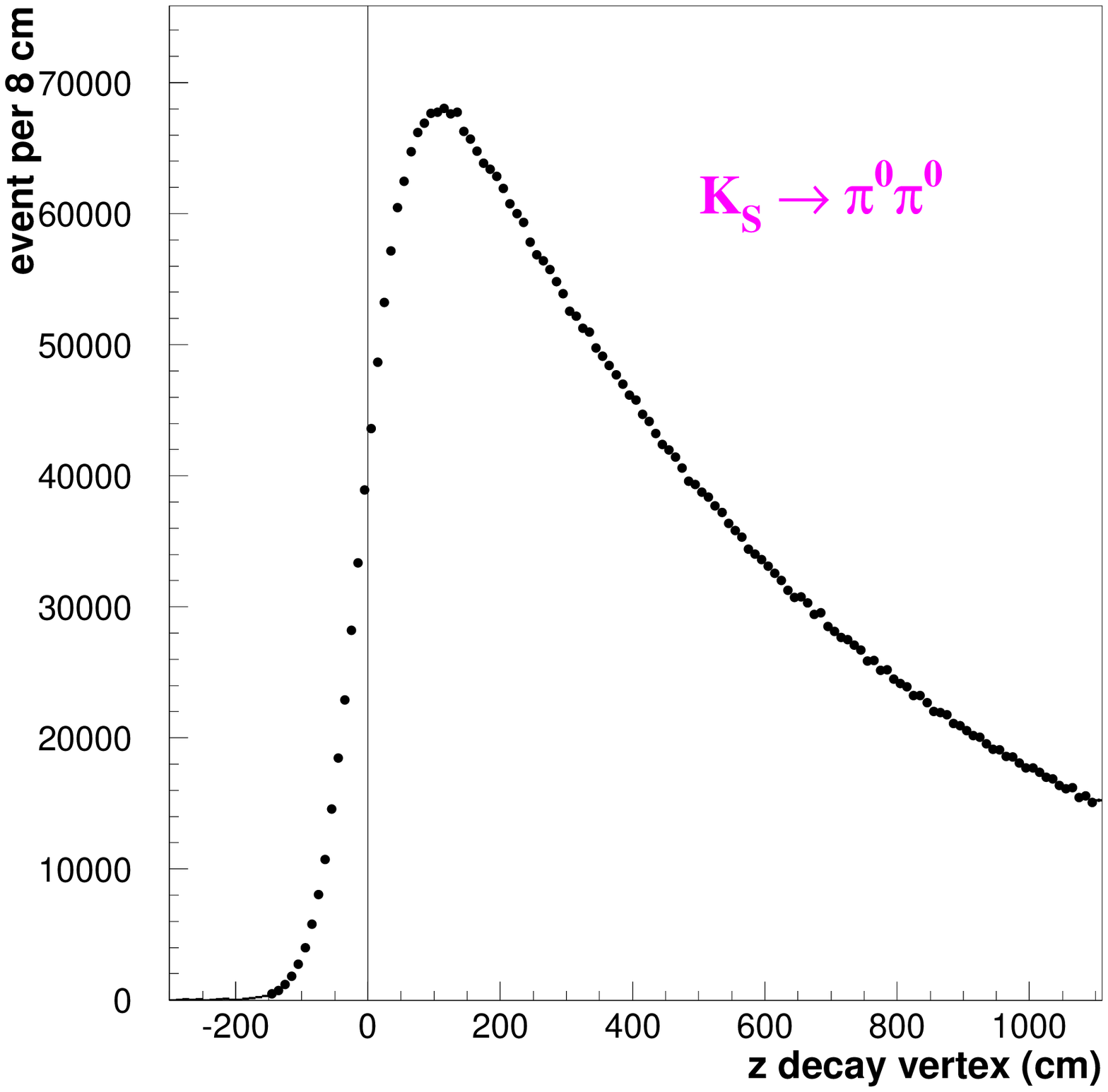,width=.5\textwidth}}
\caption{Reconstructed longitudinal vertex position for \KS neutral decays.
 An anticounter vetoes all decays occurring upstream
the fiducial decay region. The anticounter position can be obtained by 
fitting the distribution, taking into account the known detector
resolution function.}
\label{aks_decay_neutral}
\mbox{}
\end{floatingfigure}
\noindent This procedure has been cross-checked by performing all along the
data--taking some runs with a special configuration (``$\eta$'' run):
a $\pi^-$ beam is sent to 2 thin targets placed in known
positions within the fiducial region, in order to produce $\pz$ and
\ETA through charge--exchange reaction. Imposing the $\pz$ or \ETA
mass to the reconstructed 
two--$\gamma$ decays, the distance between the targets and the
calorimeter is reproduced within $3\times 10^{-4}$. This uncertainty on 
energy scale corresponds to an error of $2\times10 ^{-4}$ on the double 
ratio. The value of the \ETA mass used for this check, and the \KZ mass as well,
have been measured from the copious data collected in \KL--only and
``$\eta$'' runs during the 2000 data--taking. To this purpose the
$3\pz$ decays were used, where the vertex position can be fixed
imposing the $\pz$ mass, which is known with $4\times 10^{-6}$
accuracy~\cite{pdg2000}, to the three $2\gamma$ pairs. In this way we
can measure the ratios $M_K/M_{\pi^0}$ and $M_{\eta}/M_{\pi^0}$, which
are independent from the energy scale setting.
A sample of $128~\times~10^{6}~\kl\to3\pz$ and
$264~\times~10^{3}~\eta\to3\pz$ candidates  was selected with
negligible background. The potentially most dangerous source of  systematic
error, namely the calorimeter non--linearity, is
suppressed with a tight cut on the photon energy asymmetry: 
$0.7 < 6E_{\gamma}/E_{tot} < 1.3$ \\
 After this cut the systematic error is
dominated by other reconstruction effects, such as non--uniformity and 
uncertainty on the energy sharing among clusters. The mass distribution
for the final sample ($655\times10^{3}~\kl\to3\pz$ and
$1134~\eta\to3\pz$) is shown in figure \ref{m3pi0}. Final values
are~\cite{etak0}:
\begin{center}
$ M_{\kz}  \ = \ 497.625 \ \pm \ 0.001{\text (stat)} \ \pm \ 0.003{\text (MC)} \ \pm \  0.03
1{\text (syst)} \ \text{MeV/c}^2 $ \\
$ M_{\eta} ~~ \ = \ 547.843 \ \pm \ 0.030{\text (stat)} \ \pm \ 0.005{\text (MC)} \ \pm \  0.04
1{\text (syst)} \ \text{MeV/c}^2 $
\end{center}
\epsfigure{htb}{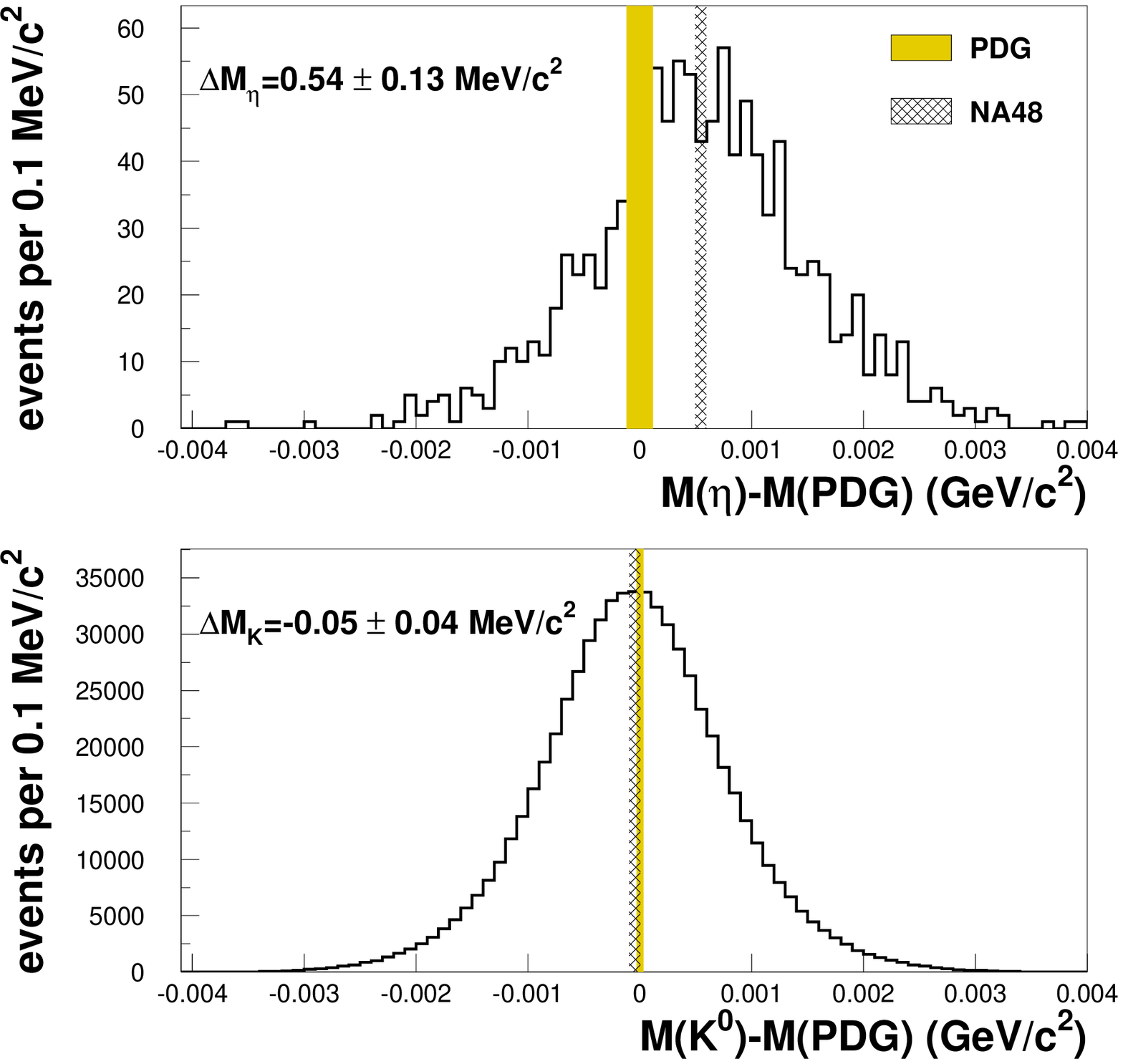}{.5}
{\mbox{Result for the \KZ and \ETA masses, compared with the
PDG values. Shaded areas show the $\pm 1\sigma$ intervals.}}
{m3pi0}

The value for the \KZ mass is in excellent agreement with the PDG 2000
world average~\cite{pdg2000} and has a similar accuracy, while for the \ETA mass
the error of this measurement is three times smaller
than the PDG one, and the agreement is poor (4.3 $\sigma$).

This result is consistent (and independent) with what observed in the
``$\eta$'' run check, where the energy scales reconstructed from
$\pz\to\gamma\gamma$ and $\eta\to\gamma\gamma$ would differ by 0.1 \%
if the PDG value of the \ETA mass was used.

\section{Measurement of \KS lifetime}
\begin{figure}[htb]
\begin{minipage}{.48\textwidth}
\centerline{\epsfig{file=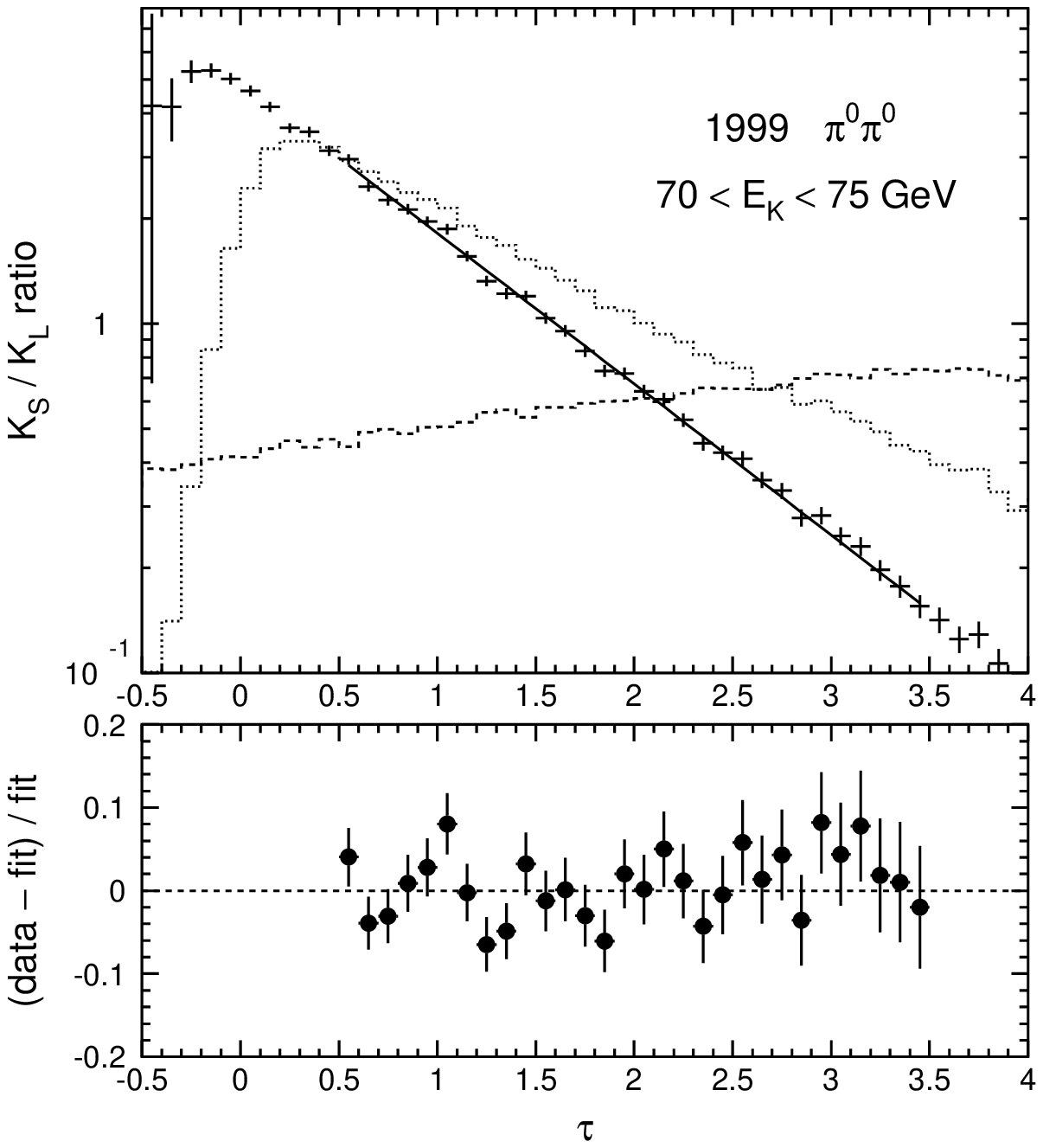,width=.75\linewidth}}
\caption{Example of fit for the \KS lifetime. The vertex distribution for \KL 
and \KS events, as well as the fitted ratio and the fit residuals are 
shown for $\pzpz$ decays in the lowest kaon energy bin (70--75 GeV).}
\label{kslifefit}
\end{minipage}  \hfil
\begin{minipage}{.48\linewidth}
\centerline{\epsfig{file=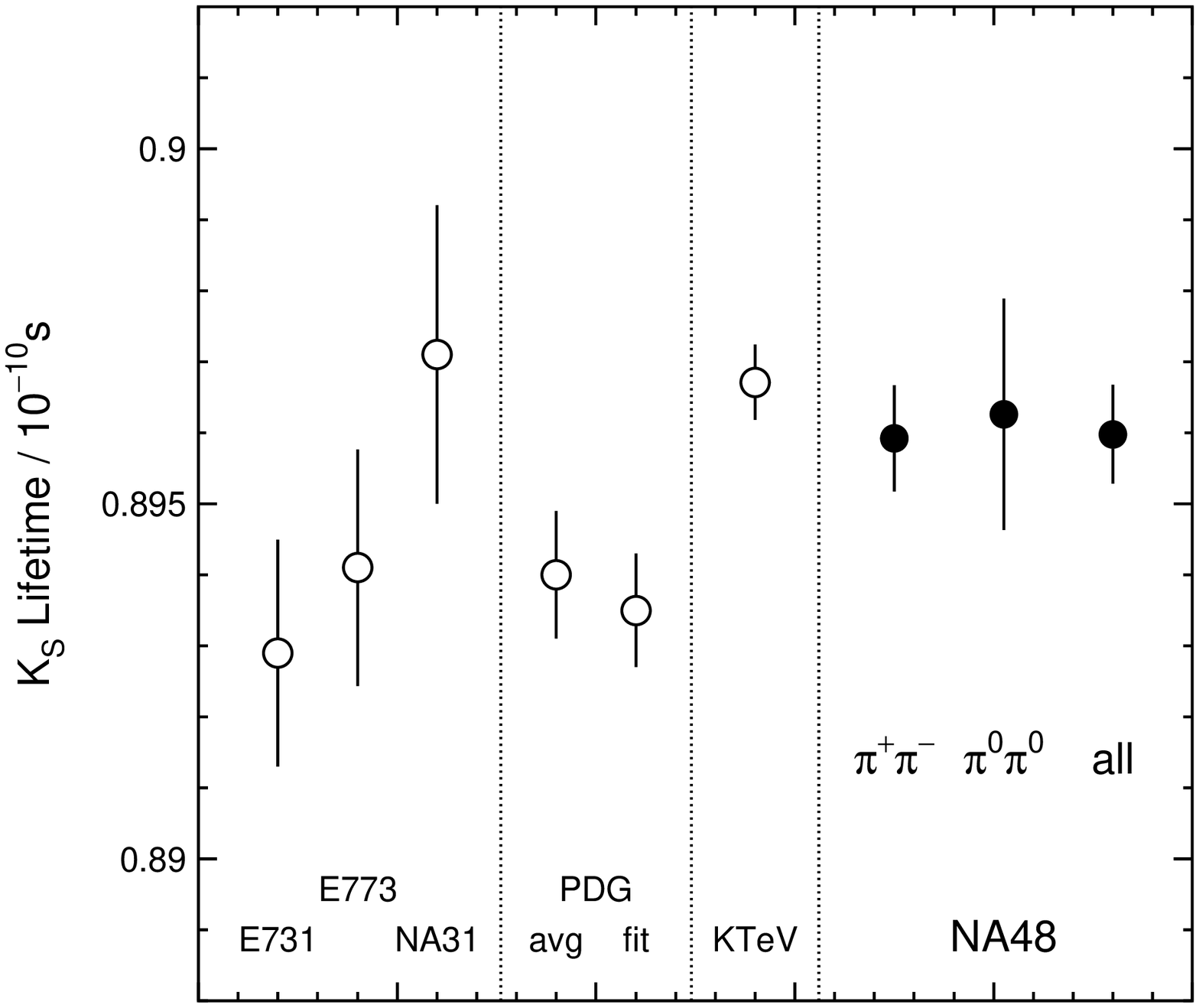,width=1\linewidth}}
\caption{Comparison between this and previous measurements of the \KS lifetime}
\label{kslife}
\end{minipage} 
\end{figure}

Another byproduct of the $\epsi'$ analysis is the precise
measurement of the \KS lifetime. The method consists in fitting the
ratio of \KS to \KL lifetime distributions in the same 20 energy bins used in
the $\epsi'$ analysis (fig. \ref{kslifefit}). In this way the \KL are
used to cancel most of 
the detector acceptance and efficiency effects. Several small
($\lesssim 3\times 10^{-4}$) residual systematic errors, essentially
the same affecting the \EPOE measurement, have been considered. The
measurement can be done independently for charged and neutral events
and consistent results~\cite{lifetime} are found: 
\begin{tabbing} 
~~~~~~~~~~~~
\= $\tau_S ~=~ (0.89592 \pm 0.00052 \text{ (stat)} \pm 0.00054 \text{(syst)})\times 10^{-10}$ s
 ~~~ \= $(\pi^+\pi^-)$ \\
\> $\tau_S ~=~ (0.89626 \pm 0.00129 \text{ (stat)} \pm 0.00100 \text{(syst)}) \times 10^{-10}$ s
     \> $(\pi^0\pi^0)$ \\
\> $\tau_S ~=~ (0.89598 \pm 0.00048 \text{ (stat)} \pm 0.00051 \text{(syst)}) \times 10^{-10}$ s
\> $(combined)$
\end{tabbing}   
in good agreement with the preliminary KTEV result~\cite{graham} and previous
measurements (see fig. \ref{kslife}).

\section{Plans for the next future}
The NA48 data can be used to study many \KL, \KS and hyperons rare
decay modes. In particular, unprecedented accuracy has been reached on 
\KS rare decays by running the experiment with a high--intensity \KS
beam configuration during the 1999 and 2000 runs. In
order to exploit this opportunity a dedicated program (NA48/1) has been
approved for the 2002 SPS run. After minor modifications of the beam
line and an upgrade of the drift chamber readout, the experiment will be able
to run with a \KS beam intensity of $2\times 10^{10}$
protons per burst, about 600 times more than the nominal intensity of
the $\epsi'$ runs. The main goals of this project are the possible
first observation of the very rare $\ks\to\pz e^+e^-$ decay (the
expected single event sensitivity is $3\times 10^{-10}$) 
and the study of indirect CP violation in $\ks\to3\pz$.

Another program, devoted to charged kaons (NA48/2), has been approved
for 2003. The beam line will be modified in order to have  a
simultaneous $K^+/K^-$ high--intensity beam and a new
beam spectrometer will be installed. The main goal is the possible
observation of direct CP violation in charged kaon decays by measuring the
$\rm K^+/K^-$ asymmetry in the $K\to3\pi$ Dalitz plot
with accuracy of the order $10^{-4}$.

\section{Conclusions}
The \EPOE program of NA48 has been completed with the successful 2001
data--taking. The new data will improve the statistical accuracy and
perform a major check of the present result: \\
\centerline{\REOE = $(15.3 \ \pm \ 2.6) \times 10^{-4}$}

New precision measurements of the \KZ and \ETA masses and
of the \KS lifetime have been obtained as a byproduct of the \EPOE
analysis.

Many other interesting results in kaon physics are being obtained from 
the collected data, and many more are expected from the future
programs, providing quantitative tests of CP violation and low--energy
hadron dynamics, highly complementary to B physics.

\section*{Acknowledgments}
We thank the organizers of the Moriond conference for the excellent
program and the very nice atmosphere. We acknowledge support from
the European Community through a Marie Curie Fellowship.

\section*{References}
\bibliography{proc}

\end{document}